\documentstyle[preprint,eqsecnum,aps]{revtex}

\begin{document}
\draft
\tightenlines

\title{Links between different models for multifragmentation}
\author{S. Das Gupta, J. Pan, I. Kvasnikova and C. Gale}
\address{Physics Department, McGill University, Montr\'eal, QC, 
H3A 2T8, Canada.}
\date{\today}
\maketitle
\begin{abstract}
We establish links between three different models for multifragmentation:
the percolation model, the lattice gas model and the statistical 
multifragmentation model.  There are remarkable similarities 
between the lattice
gas model and the statistical multifragmentation model.  For
completeness, we also
compare with a model based
upon classical molecular dynamics which gives rather different results.
\end{abstract}
\pacs{PACS numbers: 21.60.-n, 21.10.-k}
\narrowtext

\section{Introduction}

In this paper we establish relationships between three models commonly
used for the theoretical treatment of nuclear
multifragmentation.  These are (1) the percolation 
model \cite{bau88,cam88}, (2) the
lattice gas model \cite{pan1,pan2} and (3) a class of 
models that we generically call
the statistical multifragmentation model (SMFM) \cite{bon95,fai82,gro85}.  
A great deal of work
in the last approach has been done.  A complete description can be
found in Ref.~\cite{bon95} and we will essentially follow the 
SMFM model presented there. Both the percolation and statistical 
multifragmentation model have been used for years.  The lattice 
gas model is of more recent origin and can actually bridge the gap 
between those two
seemingly totally different approaches. 

The connection between the percolation and the lattice gas model was already
elaborated upon when the lattice gas model was introduced.  It is included here
for completeness.  The main objective of the present paper is to point out
that even though the lattice gas model and SMFM start off with two seemingly 
very different
premises, there are many common features between the two models.
Prompted by this similarity, we are driven to try and draw a $P-V$ 
diagram for the SMFM and to obtain the parameters of its critical 
behaviour, as for the lattice gas.  We view this current work as mostly 
theoretical and we shall make only brief references
to experimental results.  For completeness we also make some comparisons
with a classical molecular dynamics model that is in some use at the
present time.

\section{The Lattice Gas and percolation Models}

The lattice gas model has been described before \cite{pan1,pan2}.  
For completeness we
describe some of its features that are relevant for this work.

The lattice gas approach is viewed in our context as a modeling tool for
the later stages of nucleus-nucleus collisions. 
For the case of $n$ nucleons at this nuclear disassembly phase, the
lattice gas theory consists of placing those $n$ nucleons at 
$N$ lattice sites where $N\geq n$ (when $N=n$ the nucleus is at its normal
volume).  Each lattice site can be populated by one nucleon 
or none.  In the lattice 
gas model we can not squeeze the nucleus to lower than normal volume.
Later we will see that this feature is common to the SMFM also.  The ratio
$N/n$ is equal to $\rho_0/\rho_f$ where $\rho_0$ is the normal nuclear density
=0.16 fm$^{-3}$ and $\rho_f$ is the freeze out density.
The ratio $N/n$ is a parameter of the model and previously we
have found that the value 2.54 provided a good fit with data
\cite{pan1}.
Two nucleons in adjacent lattice sites will interact with each other.
We have previously taken this interaction to be attractive and equal
to $\epsilon$, irrespective of whether it is between like particles (p-p and
n-n) or between unlike particles (n-p interaction).  We shall still 
adopt this argument for the time being but we shall later improve on
this point. 
To generate an event we have to put nucleons in sites at
a given temperature, taking into account that it is energetically  
preferable to put
a nucleon adjacent to an occupied site rather than adjacent to an 
empty one. 
This Monte-Carlo sampling involves a Hamiltonian and is more involved than 
in a pure site percolation
model.  As can be readily guessed, the temperature is related 
to the excitation energy per nucleon.
Once the nucleons have been placed in lattice sites we have to specify 
the momentum of each nucleon.  The momentum of  each nucleon is generated by a
Monte-Carlo sampling of a Maxwell-Boltzmann distribution for the specified
temperature.  Thus at a given temperature we can generate an event in
which the potential energy is obtained from the number of $n-n$ bonds
and the total kinetic energy is simply the sum of kinetic energies of the
$n$ nucleons.  
We now recall how clusters are generated in the model.  Once we have put in the
nucleons in their lattice sites and generated their momenta it is easy to
calculate the cluster distribution. Two nucleons in
adjacent sites will form a cluster or be part of the same cluster if 
the relative kinetic
energy of the two nucleons is insufficient to overcome the attraction:
 $p_r^2/2\mu +\epsilon < 0$.  Here $\mu$ is the reduced mass, and
$\epsilon$ is a binding energy parameter from which the binding
energy per nucleon can be calculated.  Using this
prescription, clusters in each simulation are easily found.  It is also
clear that in general, clusters do not have to appear in their ground
state. They can be in an excited state. 
A comparison 
of yields of clusters calculated with this prescription was found in 
good agreement
with many experiments \cite{li93,bea96}.  Since the lattice gas model
has a Hamiltonian, it is possible in principle 
to obtain an equation of state. 
The $P - V$ diagram for the lattice gas model was drawn in \cite{pan2} 
for an infinite
system using the Bragg-Williams approximation.  Results from improved 
Bethe-Peierls approximation were also shown.  Clear indications of thermal
critical phenomena are seen.  It will be useful to remember here that the
$P - V$ diagram for lattice gas in the mean field approximation is similar
to that of a Van der Waals gas.  Improved calculations put the critical
temperature at 1.1275$| \epsilon |$.  The critical density is $\rho_0/2$
\cite{sti83}.

The similarities of the lattice gas model with the percolation model are 
easy to see.  For the latter we restrict ourselves to bond percolation model 
which is normally used for comparison with experiments.  In terms of
observables in heavy ion reactions, the bond percolation
model is almost identical with the lattice gas model in the limit $N=n$.
In the lattice gas model two nearest neighbours are bound to each other
if $p_r^2/2\mu+\epsilon <0$.  This probability decreases with increasing
temperature.
In the bond percolation model the probability that bonds between nearest
neighbours operate is given by a probability $p$ which similarly decreases
with increasing temperature.  Indeed it was shown that the 
phenomenological 
parametrization for $p$ used in bond percolation calculation makes the
bonding probability same in both the models \cite{pan3}.
For a certain value of $p$, a critical percolation cluster will 
appear.  
In a bond percolation model
there is a concept of temperature but the volume is not a variable.  Thus
the concept of pressure is also absent.  This is still a very valuable
model since the fragment distribution, which is the experimental 
observable, 
is dictated by percolation characteristics.  It is evident that the lattice
gas model possesses the percolation model as a subset.  In 
addition to containing all
the percolation phenomena not only at normal volume but also at larger volumes,
it also has thermal phase transitions.  The concepts
of temperature, pressure and volume are all valid concepts.  An equation of
state curve can be drawn for the lattice gas model but the percolation model
is missing a parameter and we can not thus plot a traditional equation of 
state.

\section{THE STATISTICAL MULTIFRAGMENTATION MODEL}

Our arguments follow those presented in Ref.~\cite{bon95} except 
that since we are not
as much interested in comparing with data as with connections between
different models, we will make some simplifications.  One assumes that 
equilibration has taken place in a freeze-out volume 
which is greater than normal
nuclear volume.  In this volume composites of different mass numbers appear.
The volume of any of these composites will be its normal nuclear volume, i.e.,
$k/\rho_0$ where $k$ is the mass number of the composite.  It 
then follows that if the total number of nucleons that are fragmenting is $n$,  
the excluded volume $V_{ex}$ is $n/\rho_0$ and the 
available volume for thermalization is $V_{f}=V-V_{ex}$.  The 
concept of excluded volume is also present 
in the lattice gas model.  Apart from the excluded volume the lattice gas 
model has a potential, as evidenced by the attraction $\epsilon$ between
nearest neighbour nucleons.  This results in binding energy for each cluster.
In the SMFM the specific two body interaction is not written down 
but there is clearly a potential energy and it makes its presence felt through
the explicit binding energy of clusters.  In the SMFM the 
clusters can be in excited states.  
In the lattice gas model also the clusters 
can be in excited states (excited states are not necessarily included
in the same fashion in both models but provisions are kept for
their inclusion).  Surface tension is included in the SMFM in the
binding energy relation for clusters (see below).  This appears naturally 
in the lattice gas model
as nucleons at the edge of a cluster will see an anisotropic neighbour
distribution.  It is because of interactions that the multiplicity
changes in the SMFM when either the temperature or the density varies.  
The potential energy will also change with density.  We 
therefore postulate that we have an interacting system where interactions
lead to clusters with associated binding energy.  Because of the  
short range of the interactions two different clusters do not interact except
through the excluded volume effect.  We can then apply the law of partial
pressures to obtain an equation of state.

\section{THE EQUATION OF STATE IN THE SMFM}

For a given freeze-out volume and temperature 
we can have $n_1$ nucleons, $n_2$ clusters of
size 2, $n_3$ clusters of size 3, etc\ldots  An extraordinarly 
large number of
choices are possible and the most probable distribution is obtained for
that choice of $n_1, n_2, n_3$ \ldots such that the free energy 
is minimized.  We
take this most probable distribution as the appropriate distribution for
the calculation of pressure.  Following the work of Ref.~\cite{bon95}, 
the number of clusters of size $k$ in volume $V_f$ is given by
\begin{eqnarray}
n_k=\exp\left[ k\beta\mu \right] 
\frac{V_f}{h^3}(2\pi mT)^{3/2}k^{3/2} \exp \left[ \beta F(k) \right]\ ,
\label{num}
\end{eqnarray}
where $m$=mass of a nucleon, $F(1)=0$ and for $k>1$
\begin{eqnarray}
F(k)=(W k-\sigma (T) k^{2/3})-< E^* > + T S
\end{eqnarray}
where $W = 16$ MeV is the volume energy term, 
$\sigma (T) k^{2/3}$ gives the surface
tension correction, 
$<E^* > = \pi^2T^2 / (4\epsilon_F)  $
is the average excitation energy for the composite and $S$, the 
entropy, is $S = \pi^2T / ( 2\epsilon_F )$.  For the numerical work,
we use $(4/\pi^2)\epsilon_F = 16$ MeV.  
The chemical potential $\mu$ in Eq.~(\ref{num}) 
is fixed from the condition $\sum _1^nkn_k=n$.  The pressure is calculated
from $P(\rho,T)=T/V_f (\sum n_k)$.  Here $\rho=n/(V_f+V_{ex})$.  In the above
we have used the grand canonical ensemble.  The surface tension
$\sigma (T)$ is taken as a function of temperature \cite{bon95}.  For 
the discussion to follow
the surface tension plays a crucial role.  The SMFM-as the lattice
gas type models-is
capable of producing a U shaped distribution for $Y(Z)$, the fragment
yield function, against $Z$, the fragment charge. 
This would not be possible without the surface 
tension term.  We also studied a SMFM model in which surface tension is not a
function of temperature.  The major features do not change in what we desribe
below.

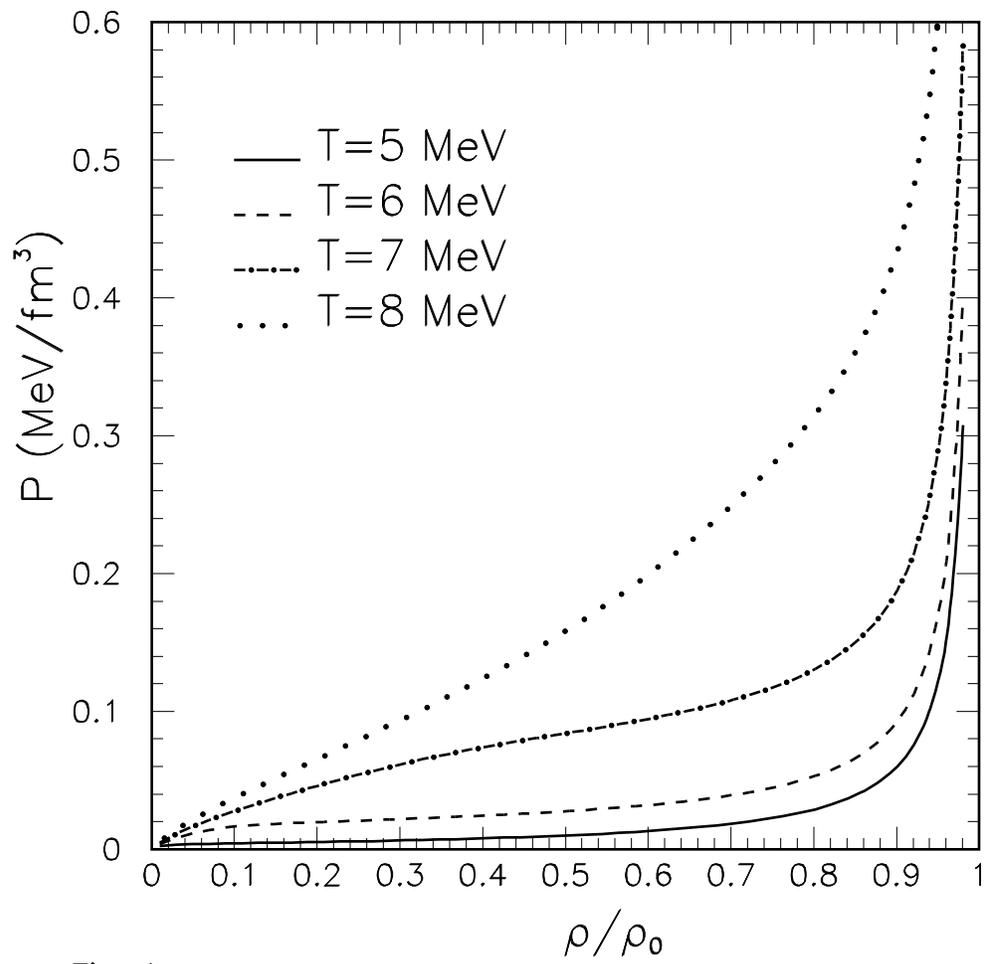
\begin{figure}
\caption{A plot of pressure against $\rho/\rho_0$ for the SMFM used in
this work.  The flat parts for the 5 and 6 MeV isotherms signify 
mixed phases.}
\label{fig1}
\end{figure}

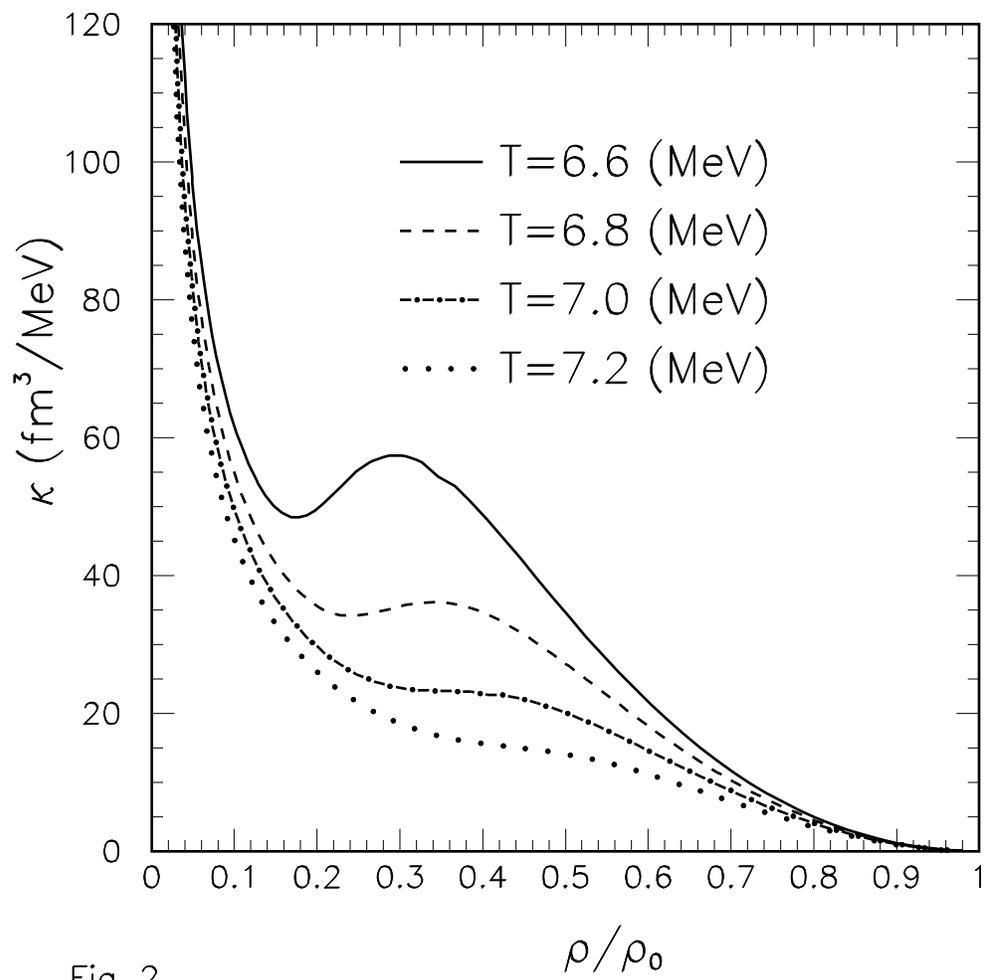
\begin{figure}
\caption{The compressibility $\kappa = 1 / \rho (\partial \rho / 
\partial P )$ at constant temperature for temperatures in the vicinity
of the critical temperature.  The maximum disappears at $T_c$.}
\label{fig2}
\end{figure}

The $P-\rho$ diagrams using this formalism were done for $A=85$ \cite{li93}, 
and also for $A=137$ (and $Z=57$), a fragment size
roughly corresponding to the projectile
fragmentation of a  gold nucleus.  In Figs.~\ref{fig1} and~\ref{fig2} 
we have shown the case of
$A=137$.  The $P-\rho$ diagram of Fig.~\ref{fig1} is strongly suggestive 
of a liquid-
gas type phase transition.  Isotherms at T = 5 and 6 MeV show regions where
pressure is very flat, 
over large variations of $\rho$.
A totally constant value of $P$ against $\rho$
in some region implies that isothermal compressibility $\kappa = 1 / \rho
(\partial \rho / \partial P )_T$ goes to infinity as will happen if one
moves along the Maxwell construction for a Van der Waals gas.  This flat
region disappears above the critical point.  Above the critical point the
compressibility is a monotonically decreasing function of density.  In 
Fig.~\ref{fig1}, the  
pressure against $\rho$ is not strictly constant in the mixed region but
has a slight rise.  Thus the compressibility is not infinite but reaches a
maximum.  This is sensible: infinities wil get replaced by 
maxima in a finite system.  We identify the critical point by the 
disappearance of the maximum in $\kappa$ (Fig.~\ref{fig2}).  
For a system of mass 137
we estimate $T_c=6.8$ MeV and $\rho_c = 0.4 \rho_0$.  For a system of 85 
particles the same criteria gave $T_c=6.2$ MeV and $\rho_c=0.14 \rho_0$.
Values of critical temperature and density for mass number 137 are not a 
great deal different from what is obtained in the lattice gas model for
an infinite system.  A binding energy of 16 MeV in nuclear matter
requires a value $\epsilon = -5.33$ MeV.  The parameters of the
critical point in the lattice gas
model will then be $T_c=6.0$ MeV and $\rho_c= \rho_0 / 2$.

\section{The equation of state in THE LATTICE GAS MODEL }

We would like to draw $P - V$ diagrams similar to above using the lattice gas 
model.  Unfortunately, this is prohibitively difficult.  While 
energy, fluctuations in energy
(essentially $C_{\rm v}$) and several other quantities can be obtained using 
Monte-Carlo simulations (later we will show results of such calculations)
calculation of pressure requires evaluation of the partition function which
can not be done using Monte-Carlo.  We can draw a $P - V$ diagram using
mean field theory (this was already done in Refs.~\cite{pan1,pan2}) but this has in
some part of the diagram a
wrong slope of pressure against $\rho$ (in fact one uses that as a signature
of phase-transition) and does not compare well with fig. 1.

While better calculations for pressure are possible, for a limited comparison
we do the following calculation.  We use a lattice of $3^3$
(this approach can not be improved by a straightforward increase of lattice
dimensions: computation with a 
$4^3$ lattice will take $2^{37}$ times as much time and is clearly beyond 
our scope)
but use periodic boundary conditions to minimize edge and finite
particle number effects
\cite{bin93}. This represents an improvement over the Bethe-Peierls
calculation done in Ref.~\cite{pan2}, as our lattice is in fact much 
larger than a Bethe-Peierls block. We use 
$P V  = T \ln Z_{gr}$, where the grand partition function 
is $Z_{gr}=\sum_0^{27}e^{\lambda n}
Z(n)$.  The value of $\lambda$ is chosen to give the required value of
$\rho/\rho_0=<n>/27$ where $<n>=\partial lnZ_{gr} / \partial \lambda$.
Finally $V=27a^3$ where $a^3=1.0/\rho_0$.  In figures 3 and 4 we have plotted
$p-V$ and $\kappa-\rho$ diagram.  Similarities with Figs.~\ref{fig1} and
\ref{fig2} are obvious.
A critical temperature between 5 and 6 MeV is suggested but one should
keep in mind the limitations induced by the 
smallness of the lattice.  The value of
$\epsilon$ used for these diagrams was $-5$ MeV.

\section{SPECIFIC HEAT}

The simple SMFM model as used here gives a maximum in $C_{\rm v}$ as one
moves across $T_c$ along a density close to the critical density. This is
shown in Fig.~\ref{fig5}.  This of course is 
in keeping with the findings of previous detailed investigations 
\cite{gro85,bon85}.  Similar maxima in $C_{\rm v}$ are found also
in the lattice gas model.  We will show our results for $n$=137 which roughly
corresponds to the projectile fragmentation of gold in the EOS TPC experiment.
The lattice sizes are taken to be $N=7^3$ and $N=8^3$ which would correspond to
two different freeze-out densities.  For this calculation we use Monte-Carlo
sampling based upon the Metropolis algorithm.  At a given temperature 
we generate
an event in which the potential energy is obtained from the number of 
nucleon-nucleon 
bonds and where the kinetic energy is simply the sum of kinetic energies of the 
nucleons.  The average of many such events gives the average energy per
nucleon, $e$, at a given temperature.  One can then obtain $C_{\rm v}$ per nucleon
by numerically differentiating $\partial e/\partial T$ or else, since the 
events
are generated in the canonical ensemble in the Metropolis 
algorithm, 
one can use 
\begin{eqnarray}
C_{\rm v} = \frac{1}{n T^2}(<E^2>-<E>^2)\ .
\label{cv}
\end{eqnarray}
Note that $<E> = e n$.

We use this opportunity to introduce an improvement to the lattice gas 
model.  In our previous work we used the same value of $\epsilon$
for attraction between like (nn and pp) and unlike (np) particles.  Although
this already gives a rather reasonable overall fit to $Y(Z)$ against $Z$
one gets a theory with di-neutrons di-protons etc\ldots  This can 
be avoided by postulating two kinds of bonds: that between the proton
and the neutron which we denote as $\epsilon_{np}$ and is attractive, and that
between identical particles (neutron-neutron and proton-proton) which we denote
by $\epsilon_{nn}$ and which is either zero or repulsive.  In accordance with
a molecular dynamics potential \cite{len90} which we will refer to 
in the next section
we have used a slightly repulsive interaction for $\epsilon_{nn}=$1 MeV.  The
attractive interaction is set at $\epsilon_{np}=-5$MeV

In each Monte-Carlo simulation we also obtain the cluster decomposition.
After a large number of simulation one obtains  $Y(Z)$ against $Z$.  We expect
that near the critical point a power law will emerge : $Y(Z)\propto Z^{-\tau}$.
The exponent $\tau$ will be minimum at the critical point.  We deduce an
effective value of $\tau$ from our simulation even when far from the critical 
point by using the formula \cite{pra95}:
\begin{eqnarray}
\frac{\sum_2^{10}ZY(Z)}{\sum_2^{10}Y(Z)}=\frac{\sum_2^{10}ZZ^{-\tau}}
{\sum_2^{10}Z^{-\tau}}
\end{eqnarray}
The variation of $\tau$ against the temperature is shown in 
Fig.~\ref{fig6}.  The 
calculated $C_{\rm v}$ against temperature is shown in 
Fig.~\ref{fig7}.  We find that the
maximum in the value of the specific heat happens at the same 
temperature where the
value of $\tau$ minimizes.  In so far as the $C_{\rm v}$ maximizes 
at the critical
temperature, the prediction of the lattice gas model is similar to that of
SMFM.  We also note that using different bonds between like and unlike 
particles (which is more realistic than the simpler case) actually enhances
the peak.

In detail the calculated values of $C_{\rm v}$'s 
differ, specially at low temperature.
Here the SMFM is clearly more realistic.  At low temperature, the breakup
event will have one large cluster and the specific heat of this cluster will
dominate the net contribution.  In SMFM the low energy excitation energy 
characteristics is put in by hand and hence $C_{\rm v}$ has been forced to go
like $T$.  In the lattice gas model the low temperature value of the
specific heat per particle will simply be 3/2.  This comes from the classical 
simulation of the kinetic
energy of the constituent nucleons.  Stated another way, the excited states
of each cluster is put in differently in the two models.

One can use the lattice gas model for predicting a caloric curve
(temperature plotted against excitation energy), provided
one assumes that calculations using the lattice gas model are valid beyond
a certain temperature only.  The uncertainty below this
temperature can be absorbed in one constant.  In 
Fig.~\ref{fig7}  we
show a caloric curve obtained in the lattice gas model.  In the figure we
also show a comparison with experimental data.  To produce the caloric curve
we use
\begin{eqnarray}
e(T)=\int_0^tC_{\rm v} dT'+\int_t^TC_{\rm v} dT'= \varepsilon + 
\int_t^TC_{\rm  v}  dT'
\end{eqnarray}
We ignore the prediction of the lattice gas model below a temperature $t$
(taken in this work to be 3.2 MeV) and we thus introduce a 
parameter $\varepsilon$.  The
caloric curve beyond the temperature 3.2 MeV is shown in Fig.~\ref{fig8}.  
In the ALADIN  
results \cite{poc95}, $T$ against the excitation energy is considerably 
flatter at 
temperature around 5 MeV, 
implying a large increase in the value of the specific heat at 
this temperature.  We
do not get such a large value of $C_{\rm v}$ in our calculations.  The
appearance of a 
possible plateau in the caloric curve 
is also less dramatic in the recent EOS TPC data \cite{hir96}.

We end this section by summarizing the comparison between the lattice gas model
and SMFM.  As we have shown there are many common features in the two models.
The lattice gas model has the advantage of having considerable
recognition in statistical physics.  It defines a two body interaction 
which is used
for all its predictions.  Phase transitions in the model are well-studied
and well-defined.
Another advantage is that fragment formation in non-spherical geometries
can
be studied.  Such geometries do occur in heavy-ion collisions
\cite{pha93}.  The advantage of SMFM type models is that binding energy, 
surface
tension and excited states can all be inputs in the model.  
This makes
accurate comparison with experiments more feasible.  No detailed theory of 
phase transitions in this type of models has yet been completed.  It is not 
clear how non-spherical geometries might influence the results.

For completeness, we now turn to a discussion of how classical 
molecular dynamics predictions differ from those of the previous
approaches. 

\section{Nuclear properties in classical molecular dynamics}

Here, we choose to study equilibrium properties of nuclear systems 
using the
methods of classical molecular dynamics. Our first task is to specify
the nucleon-nucleon interaction. We use a
neutron-proton interaction, $v_{\rm n p} ( {\bf r})$, which is attractive
for large values of the separation distance ${\bf r}$, but repulsive for
small values.  The like-particle nuclear potentials, $v_{\rm n n}$ and 
$v_{\rm p p}$ are taken to be identical and purely repulsive. Those features
are chosen to satisfy the basic requirements of nuclear phenomenology. 
Our interparticle nuclear potentials are given as
a combination of Yukawa interactions \cite{len90}:
\begin{eqnarray}
v_{\rm n n} ( {\bf r} < {\bf r}_c )\ &=&\ v_0\ \left[ \exp ( - \mu_0 {\bf
r}) / {\bf r}\ -\ \exp (- \mu_0 {\bf r}_c ) / {\bf r}_c \right]\ , \\
v_{\rm n p} ( {\bf r} < {\bf r}_c )\ &=&\ v_r\ \left[ \exp ( - \mu_r {\bf
r}) / {\bf r}\ -\ \exp (- \mu_r {\bf r}_c ) / 
{\bf r}_c \right]\nonumber \\
          & &-\ v_a\ \left[ \exp ( - \mu_a {\bf
r}) / {\bf r}\ -\ \exp (- \mu_a {\bf r}_c ) / {\bf r}_c \right]\ .
\label{pot}
\end{eqnarray}
In the above, ${\bf r}_c$ is fixed at  5.4 fm, a
simple and adequate cutoff. It is also implied that 
\begin{eqnarray}
v_{\rm n n} ( {\bf r} \geq {\bf r}_c )\ =\ v_{\rm p p} 
( {\bf r} \geq {\bf r}_c )\ =
0\ .
\end{eqnarray}
The parameters of the potentials are $v_0$ = 373.118 MeV fm, $v_r$ =
3088.118 MeV fm, $v_a$ = 2666.647 MeV fm, $\mu_0$ = 1.5 fm$^{-1}$, $\mu_r$
= 1.7468 fm$^{-1}$, and $\mu_a$ = 1.6 fm$^{-1}$. The nuclear 
compressibility coefficient, $K$,
is 250 MeV for this choice of parameters \cite{len90}.   In the case of
proton-proton interactions, the Coulomb interaction can be added
separately. However, in order to make a consistent comparison, 
the Coulomb effects were left out in this work.

In the molecular dynamics simulation, the particles are propagated
in phase space by integrating Newton's equations of motion through 
a ``leap-frog'' algorithm \cite{all87}. Because we shall address shortly
the issue of specific heat, we chose to perform simulation of canonical
ensembles. One can show \cite{all87} that constant temperature dynamics
are generated by the following equations of motion for the position ${\bf
r}_i$ and the momentum $ {\bf p}_i$ of each individual particle $i$ of
mass $m$ subject to a force ${\bf f}_i$:
\begin{eqnarray}
\dot{{\bf r}}_i\ &=&\ {\bf p}_i / m \\
\dot{{\bf p}}_i\ &=&\ {\bf f}_i\ -\ \xi  {\bf p}_i
\end{eqnarray}
The quantity $\xi$ acts as a ``friction coefficient'' that is
recalculated at each time step and keeps the temperature constant to a
high level of precision \cite{iou96}. The system is initialized in
momentum space from a
Monte Carlo sampling of a Boltzmann distribution, and then cooled to the
desired temperature. This method ensures a proper thermodynamic
equilibrium distribution for the nucleons in interaction. The nucleons
are confined to box whose volume is adjusted to yield the desired
density. 

In our canonical ensemble simulations, the pressure can then be
calculated for a system of $n$ interacting nucleons by means of a virial
expansion \cite{wal85}:
\begin{eqnarray}
\frac{P_{virial}}{\rho  T}=1-\frac{1}{6 n  T}\left
\langle\,
\sum_{i} \, \sum_{j>i} r_{ij}\,\frac{\delta v_{ij}}{\delta r_{ij}}\, 
\right\rangle \ .
\end{eqnarray}
Note that we are using units where the Boltzmann constant, $k_B$, is
unity. The potential energy between two nucleons $i$ and $j$, separated
by a distance $r_{i j}$,  is written
as $v_{i j}$. It is straightforward to calculate the average total energy and
the average total energy squared, so that the specific heat, $C_{\rm v}$, can
easily inferred from Eq.~(\ref{cv}). 
Note that microcanical molecular dynamics has been used in the 
recent past to study the multifragmentation behaviour of heavy ion
systems \cite{bon94,pra95}.

On Fig.~\ref{fig9} we plot pressure isotherms obtained in molecular
dynamics of our nuclear fluid with the potential of Eq.~(\ref{pot}), for
a system of $A=137$ and $Z = 57$. We
restrict our calculation to the range of temperatures and densities
related to this comparative study with the lattice gas model and SMFM.  In
this temperature range, one realizes that
the molecular dynamics predictions are devoid of the plateau 
regions characteristic of the above two approaches. It appears that the
critical behaviour of the molecular dynamics ensembles is in
fact at a temperature just below 1 MeV and at nuclear densities
around  $ \approx \rho_0 / 3$. Those parameters follow from our own simulations
\cite{iou96} and are also in line with those found in a similar 
effort, using however  a slightly different potential \cite{pra95}.
Turning to the calculations  of $C_{\rm v}$ plotted on Fig~\ref{fig10}, we come 
to conclusions similar to the ones reached in our study of the equation
of state,
namely that the specific heat does not exhibit any striking behaviour.
There are no noticeable peaks that
would follow from a flattening of the caloric curve, at the temperatures
and densities dictated by our comparative study. At high
temperatures, $C_{\rm v}$ tends to its value for noninteracting gases:
$C_{\rm v}$ = 3/2.  We will not insist further 
here on a detailed molecular
dynamics study of
the thermodynamic properties of the nuclear fluid at finite temperatures,
as this does not pertain directly to the work at hand. This is however
done and will appear elsewhere \cite{iou96}. 

\section{Conclusions}

We have shown that some popular theoretical approaches to nuclear
multifragmentation shared several common features. Those models are the
percolation, lattice gas and statistical multifragmentation.  The connection
between the lattice gas and percolation models is more apparent and has
been pointed out before, but
two models that previously appeared rather unrelated, the lattice gas and SMFM,
have some remarkable similarities to each other. We see this point as
being quite important, as it represents a rather satisfying unity aspect from 
a theoretical perspective. This is the aspect we chose to insist on in
this work.

We also have shown some comparisons with
calculations done in a classical molecular dynamics scenario. The
results there are quite different. There are of course obvious deficiencies 
associated with the use of a classical approach to a system in a regime
where the quantum aspects are undoubtedly important. 
We plan to investigate those deficiencies, and their
effects, further.

\section{acknowledgements}
Our research is supported in part by the Natural Sciences and
Engineering Research Council of Canada and in part by the Fonds FCAR of
the Qu\'ebec Government.

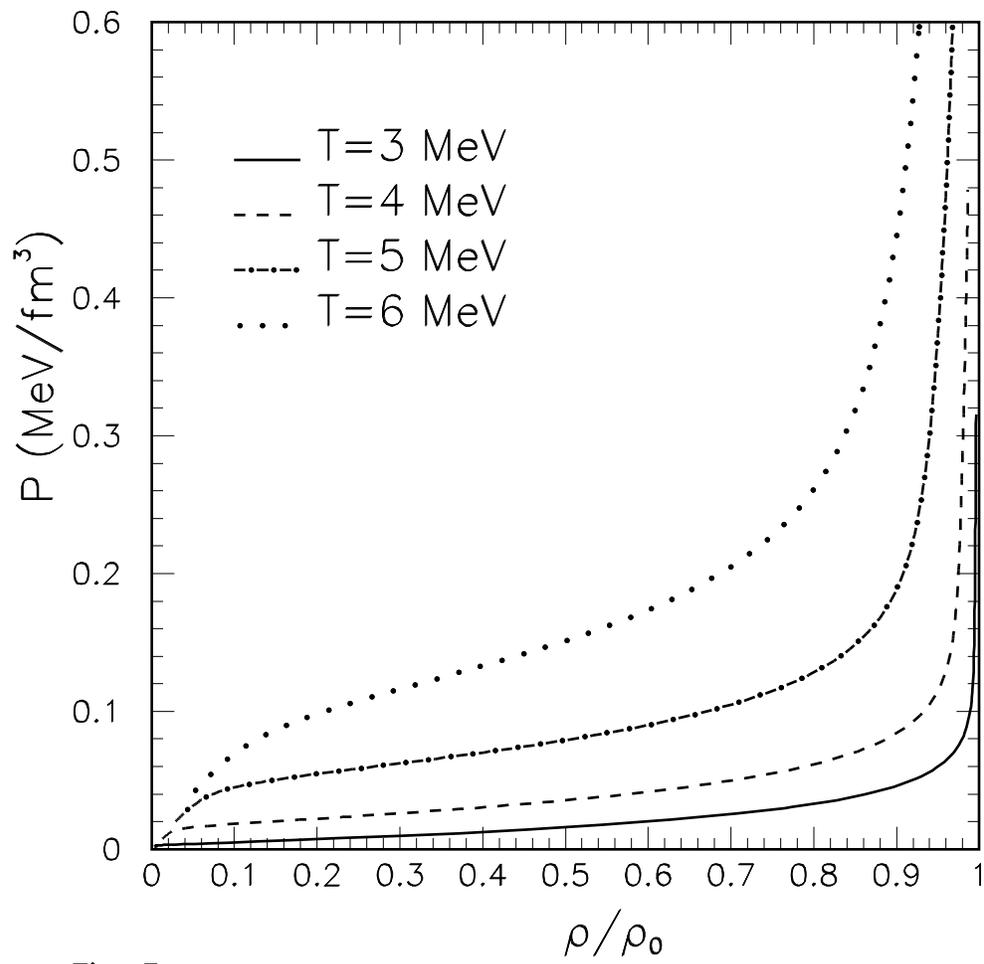
\begin{figure}
\caption{The $P-\rho$ diagram from the lattice gas model obtained by
using a $3^3$ lattice and imposing periodic boundary conditions.}
\label{fig3}
\end{figure}

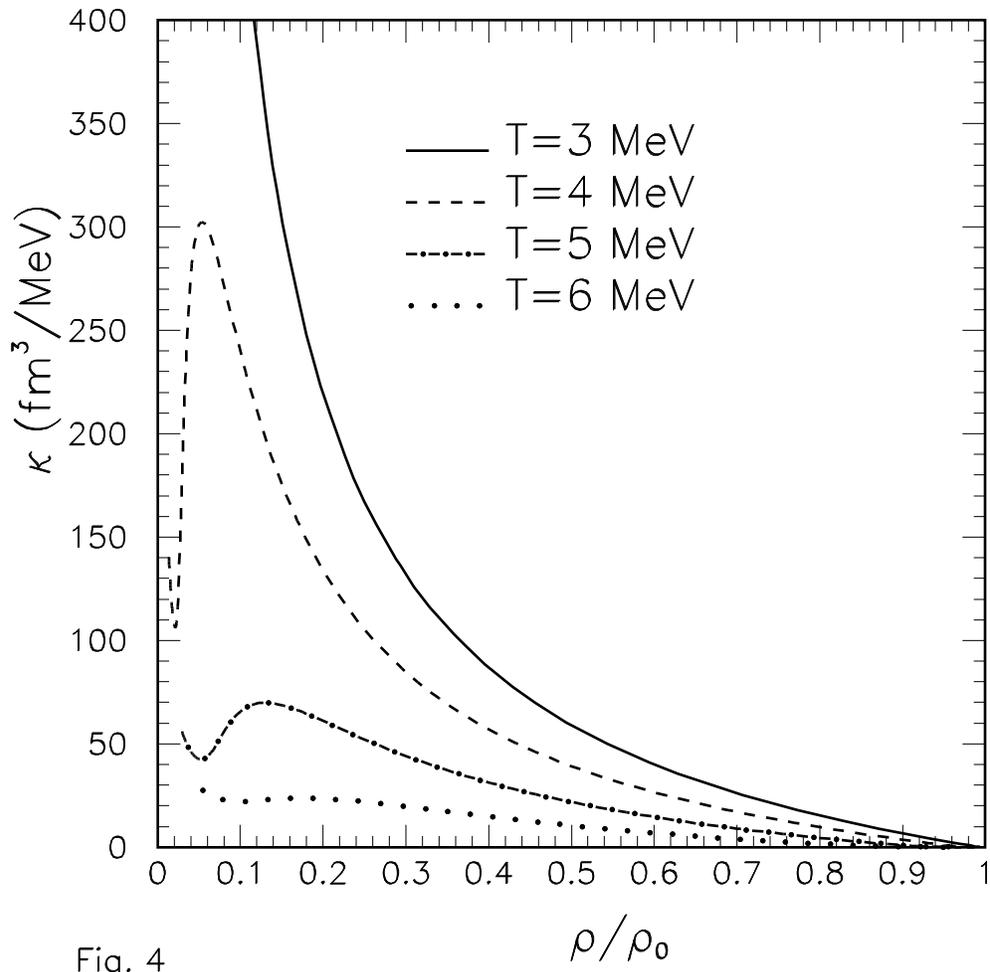
\begin{figure}
\caption{The isothermal compressibility $\kappa$ at various 
temperatures for the $3^3$ lattice.}
\label{fig4}
\end{figure}

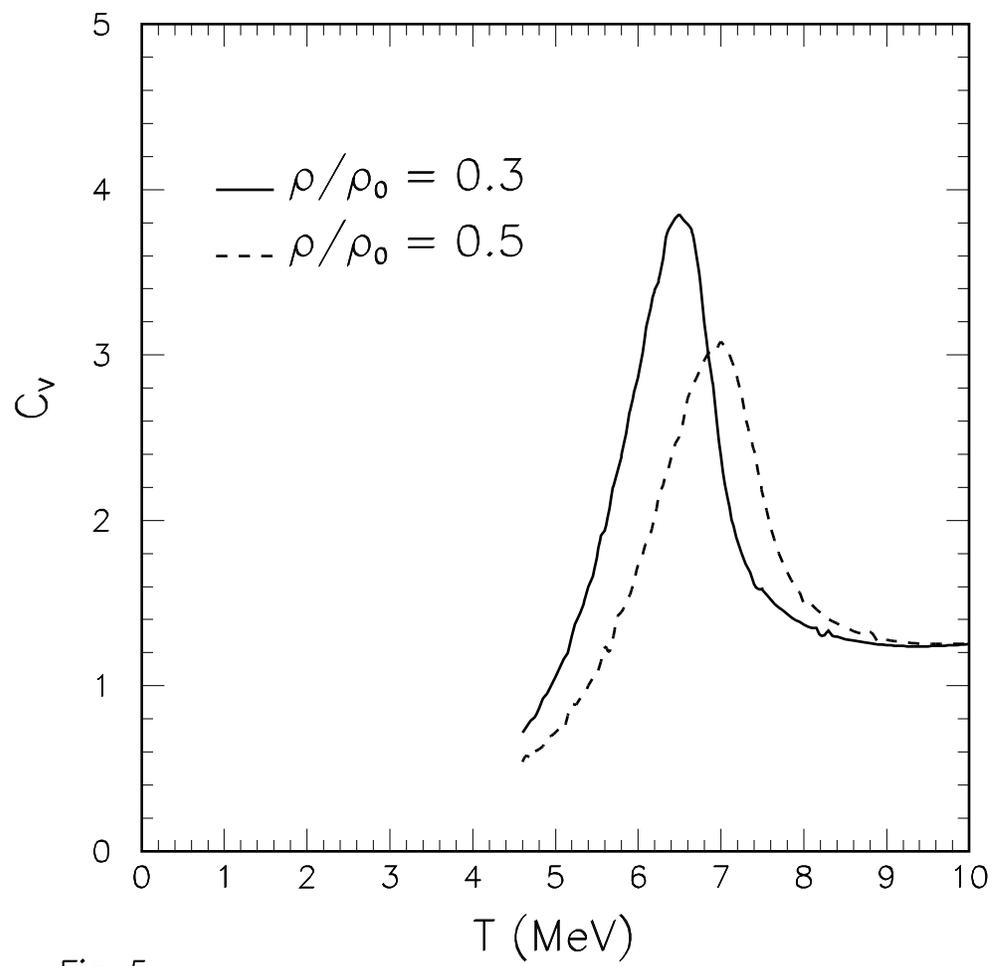
\begin{figure}
\caption{The specific heat per particle for the SMFM used in the
text for a system of 137 nucleons.  Two different freeze-out 
densities are used.
The specific heat is dimensionless here as the unit for both heat and 
temperature is MeV.}
\label{fig5}
\end{figure}

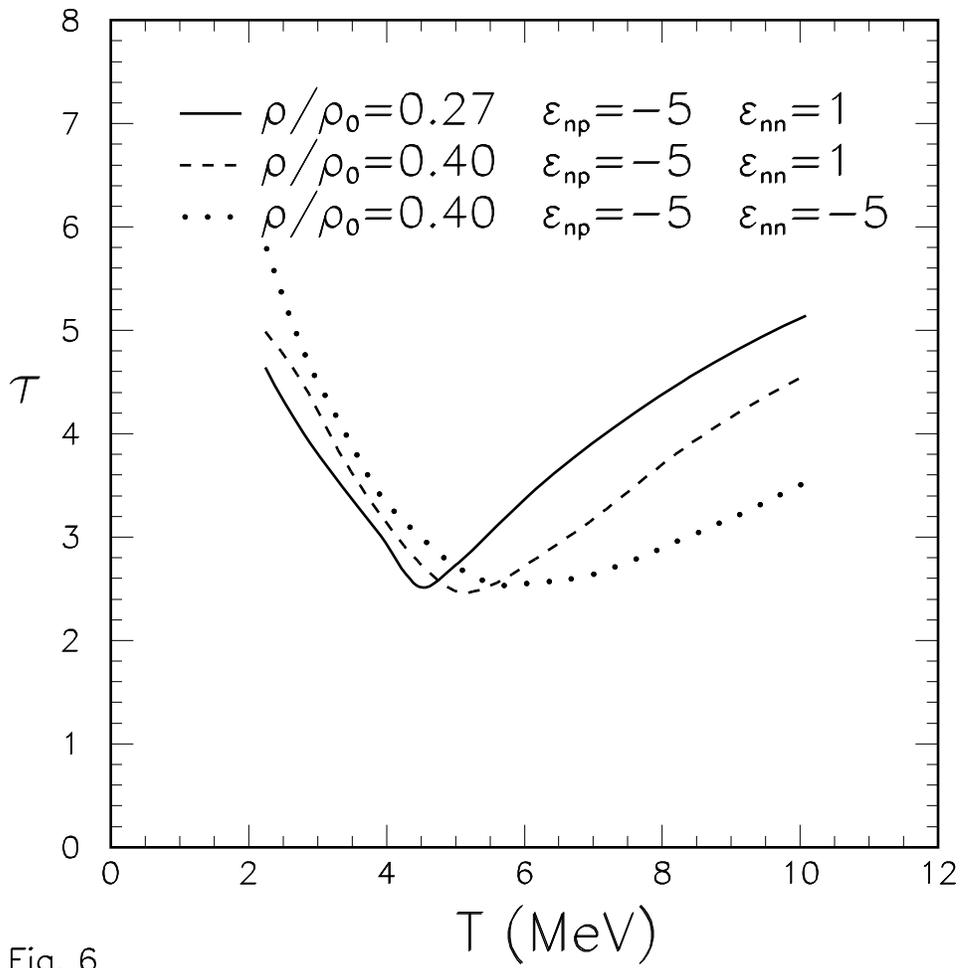
\begin{figure}
\caption{The exponent $\tau$ calculated in the lattice gas model for
a system of 137 particles.  We show results when the same value of $\epsilon$
is used for like and unlike particles and also when they are different.
In the latter case we show results for two densities.}
\label{fig6}
\end{figure}

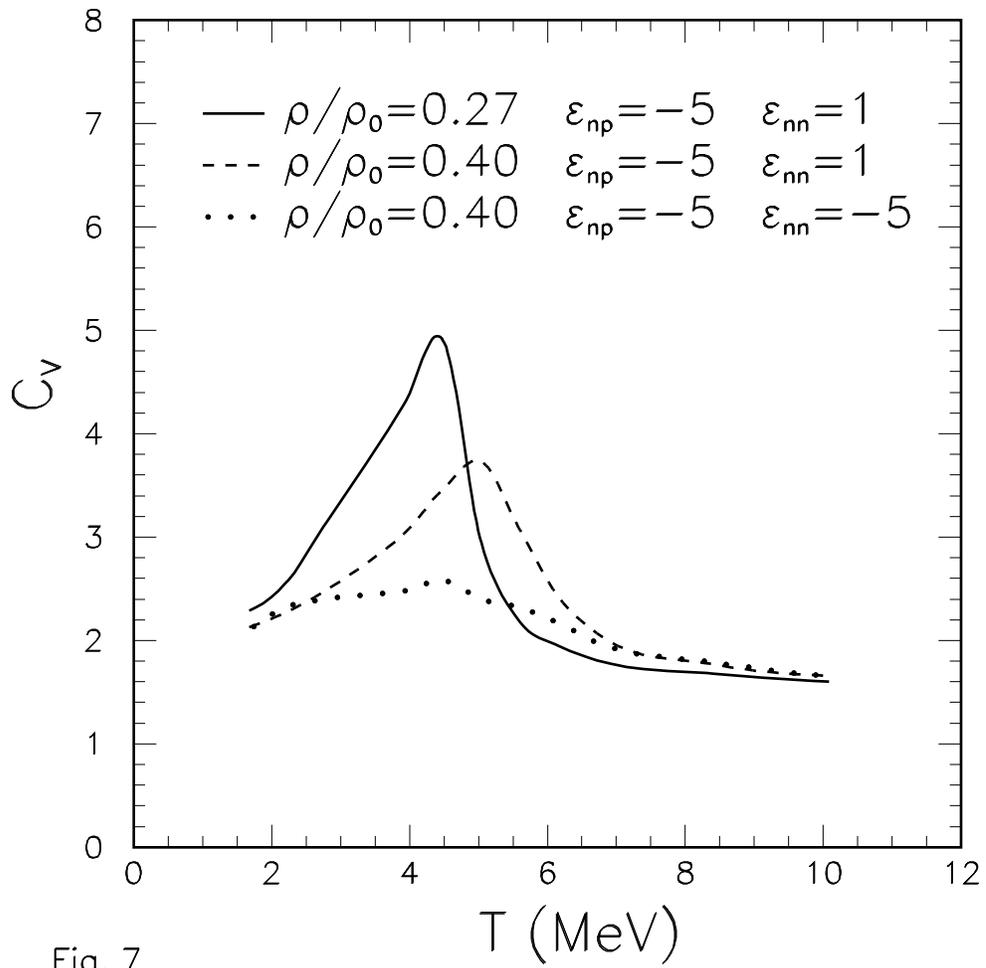
\begin{figure}
\caption{The specific heat per particle in the lattice gas model for
a system of 137 particles.  The maxima here closely coincide with the
positions of the minima in Fig. 6.}
\label{fig7}
\end{figure}

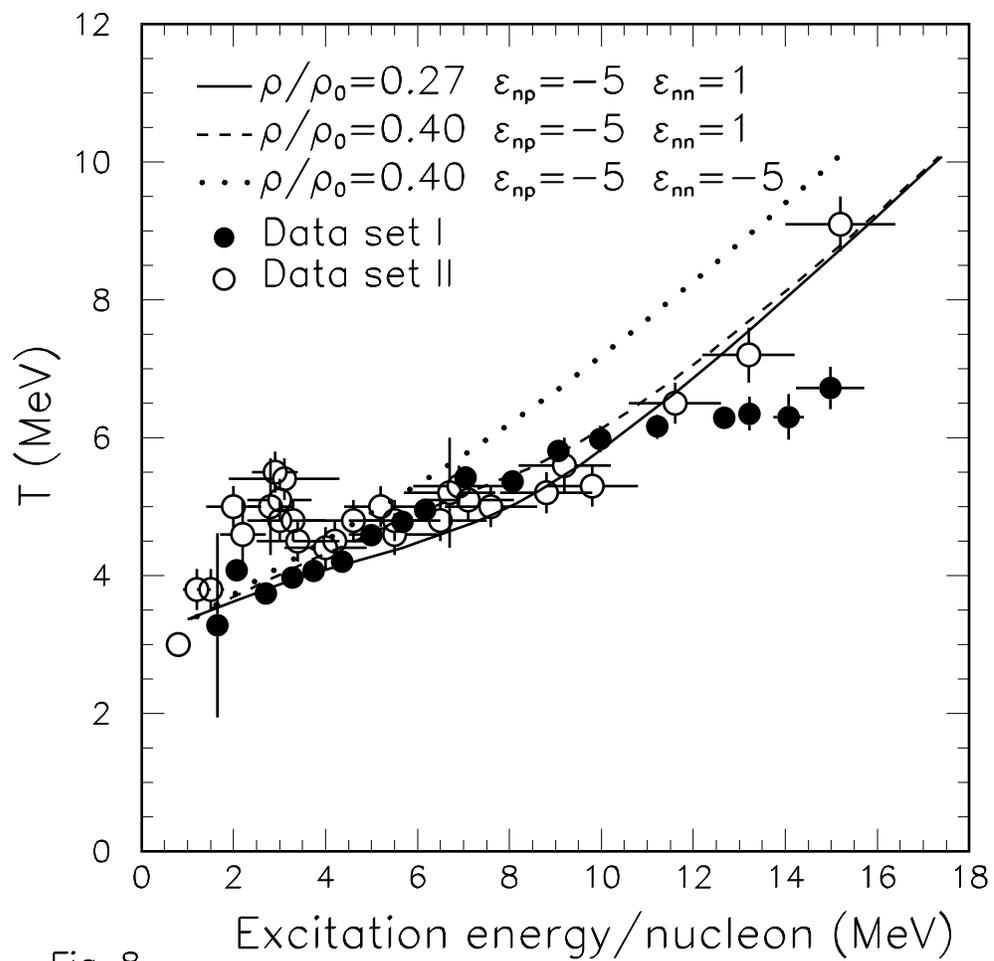
\begin{figure}
\caption{The caloric curve calculated according to lattice gas model
for a system of 137 particles and compared with ALADIN data (data set II),
Ref.~\protect\cite{poc95} and with EOS TPC data (data set I), 
Ref.~\protect\cite{hir96}.}
\label{fig8}
\end{figure}

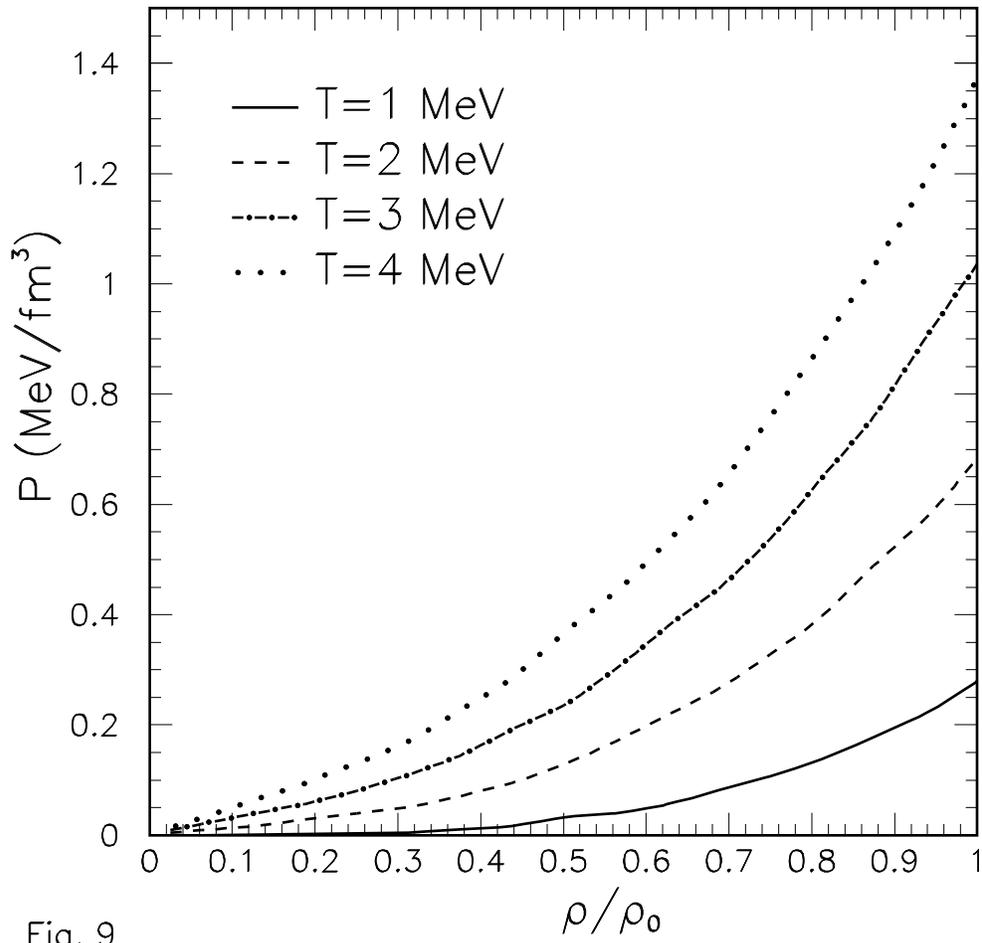
\begin{figure}
\caption{Pressure isotherms calculated in classical molecular 
dynamics, for temperatures ranging from 1 to 4
MeV. We limit ourselves to densities lower than the equilibrium nuclear matter
density, $\rho_0$. The error bars are not shown. The errors are
negligible below $\rho / \rho_0$ $\approx$ 0.5, and slowly rise to be
$\approx$ 5\% near $\rho / \rho_0$ = 1.  }

\label{fig9}
\end{figure}

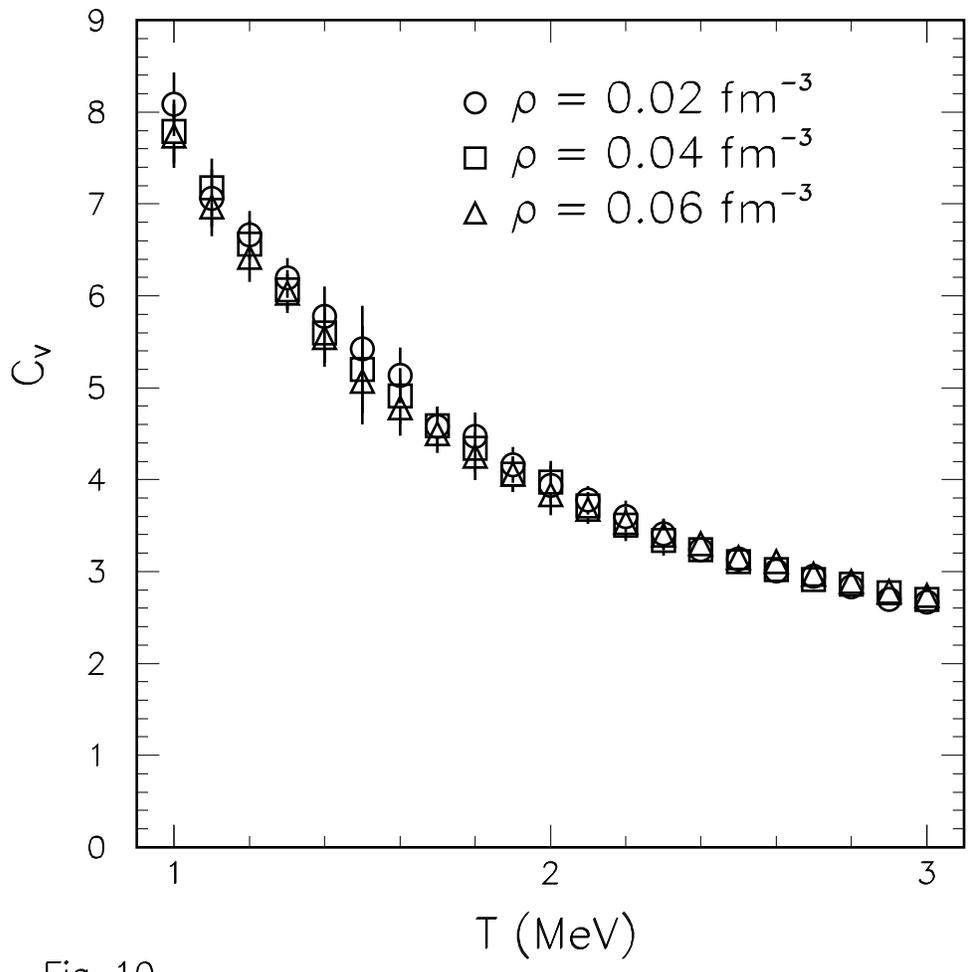
\begin{figure}
\caption{The specific heat per nucleon,  $C_{\rm v}$, calculated in
classical molecular dynamics. Three representative values of nuclear
density are shown. The errors are statistical.}

\label{fig10}
\end{figure}

\end{document}